\documentstyle[12pt]{article}

\renewcommand{\thefootnote}{\fnsymbol{footnote}}
\renewcommand{\r}[1]{(\ref{#1})}
\newcommand{\R}{{\sf R\hspace*{-0.9ex}\rule{0.15ex}%
{1.5ex}\hspace*{0.9ex}}}

\begin{document}
\thispagestyle{empty}
\newlength{\defaultparindent}
\setlength{\defaultparindent}{\parindent}

\begin{center}
{\large{\bf SO(3) GAUGED SOLITON OF AN 0(4) SIGMA MODEL ON $R_3$}} 
\vspace{0.5cm}\\

{\large K. Arthur}\\
{\it School of Mathematical Sciences,}\
{\it Dublin City University}\\
{\it Glasnevin, Dublin 9, Ireland.}\vspace{0.5cm}\\

{\large D.H. Tchrakian\footnote{Supported in part by CEC 
under grant HCM--ERBCHRXCT930362},}\\
{\it Department of Mathematical Physics,}\
{\it St Patrick's College Maynooth,}\
{\it Maynooth, Ireland}\\
{\it School of Theoretical Physics,}\
{\it Dublin Institute for Advanced Studies,}\\
{\it 10 Burlington Road,}\
{\it Dublin 4, Ireland.}\vspace{0.5cm}

\end{center}

\bigskip
\bigskip
\bigskip
\bigskip
\begin{abstract} Vector $SO(3)$ gauged $O(4)$ sigma models on $\R_3$ are 
presented. The topological charge supplying the lower bound on the 
energy and rendering the soliton stable coincides with the Baryon number 
of the Skyrmion. These solitons have vanishing magnetic monopole flux. To 
exhibit the existence of such solitons, the equations of motion of one of 
these models is integrated numerically. The structure of the conserved 
Baryon current is briefly discussed.
\end{abstract}

\vfill
\setcounter{page}0
\renewcommand{\thefootnote}{\arabic{footnote}}
\setcounter{footnote}0
\newpage

\newcommand{\ra}{\rightarrow}

\newcommand{\dd}{\mbox{d}}
\newcommand{\ee}{\end{equation}}
\newcommand{\be}{\begin{equation}}
\newcommand{\ii}{\mbox{ii}}
\newcommand{\pa}{\partial}
\newcommand{\vep}{\varepsilon}
\newcommand{\bfR}{{\bf R}}
\newcommand{\lm}{\lambda}

\pagestyle{plain}

\section{\bf Introduction}
\setcounter{equation}{0}

\parskip0.5truecm

The problem of gauging a sigma model has been considered in the past, 
first by Fadde'ev\cite{Fadde'ev} and also in the works of 
Witten\cite{Witten} and Rubakov\cite{Rubakov} for the 
Skyrme\cite{Skyrme} model on $\R_3$, which is esentially the $O(4)$ 
sigma model. Rubakov\cite{Rubakov} in particular considers the 
properties of the soliton in the $SU(2)$ gauged Skyrme model. The purpose 
of the present work is to construct a topologically stable soliton in a 
particular $SO(3)$ gauged Skyrme model.

The problem of gauging a Sigma model such that the resulting system 
supports topologically stable finite action or energy solutions was recently 
considered in \cite{Arthur} and \cite{Manvelyan}, the first for the $U(1)$ 
gauged $CP^1$ Grassmanian model on $\R_2$ and the second for $SO(2n)$ 
gauged Grassmanian models on $\R_{2n}$.

The only limitation of these models is that their definitions are restricted to 
{\it even} dimensions and the physically important problem of the soliton 
of the Skyrme model\cite{Skyrme}, which is the $O(4)$ Sigma model
defined on $\R_3$. Indeed with this problem in mind, a peculiar gauging of 
the $O(3)$ model on $\R_2$ was proposed in \cite{Arthur}. Denoting the 
Sigma model fields $\phi^a =(\phi^{\alpha} ,\phi^3)$ with $\alpha =1,2$, 
and $\phi^a \phi^a =1$, the gauging prescription proposed was stated via 
the minimal coupling in terms of the covariant derivative $D_i \phi^a =(D_i 
\phi^{\alpha} , D_i \phi^3 )$ as:

\begin{equation}
\label{1.1}
D_i \phi^{\alpha} =\partial_i \phi^{\alpha} +A_i\varepsilon^{\alpha \beta} 
\phi^{\beta},\qquad \qquad  D_i\phi^3 =\partial_i \phi^3.
\end{equation}

The topological invariants we sought to employ in \cite{Arthur} were the 
integrals of densities which were total divergences. Since the topological 
charge density of the $O(3)$ sigma model is not a total divergence but is 
only {\it locally} a total divergence\cite{Zakrzewski}, we ignored the 
$O(3)$ model gauged with the $U(1)$ field according to \r{1.1} and 
proceded\cite{Arthur} instead to gauge the $CP^1$ model. Subsequently 
however the $O(3)$ model on $\R_2$, gauged according to \r{1.1} was 
shown by Schroers\cite{Schroers} to support topologically stable and even 
self-dual solutions.

It is our aim here to gauge the $O(4)$ gauge model\cite{Skyrme} on $\R_3$ 
employing an extended version of the minimal coupling prescription 
given by \r{1.1}, where the $U(1)$ is replaced by the $SO(3)$ gauge group, 
and to show that the corresponding gauged system supports stable solitons. 
While we shall restrict our considerations here to this $3$ dimensional 
case, it should be noted that all our considerations in the present work can 
be extended to the $d$ dimensional case systematically, with the $SO(d)$ 
gauging of the $O(d+1)$ Skyrme-Sigma model\cite{Roche}. The starting 
point in that case would be the generalisation of the minimal coupling 
prescription \r{1.1} to

\begin{equation}
\label{1.2}
D_i\phi^{\alpha} =\partial_i \phi^{\alpha}+ (T^{\alpha})^{\beta \gamma} 
A^{\beta}_i \phi^{\gamma},\qquad \qquad D_i\phi^{d+1} =\partial_i 
\phi^{d+1}
\end{equation}

\noindent
where $(T)$ are the generators of $SO(d)$ in the {\it vector} 
representation, with $\alpha =1,2,..,d$, and $\phi^a \phi^a =1$. In what 
follows, we shall restrict ourselves to the case of $d=3$. In that case, the 
minimal coupling given by \r{1.2} corresponds to gauging the vector 
$SU(2)$ subgroup of the chiral $SU(2) \times SU(2) \sim O(4)$.

Note that the above gauging \r{1.2} for $d=3$ differs from that employed in 
\cite{Rubakov}. This is easily seen by identifying the $SU(2)$ group valued 
field $U$ as $U=\phi^a \tau^a , U^{-1} =\phi^a \tilde \tau^a$, with $\tau^a 
=(i\sigma^{\alpha} ,1)$, $\tilde \tau^a =(-i\sigma^{\alpha} ,1)$, and $A_i = 
-iA^{\alpha}_i \sigma^{\alpha}$. In components then, it turns out that 
$(D_i U)^{\alpha}=D_i \phi^{\alpha} -A^{\alpha}_i \phi^4$ and $(D_i U)^4 
=D_i\phi^4 +A^{\alpha}_i \phi^{\alpha}$, in the notation of \r{1.2}.

\section{\bf The topological Invariant and the Model}

The cornerstone in our construction of the gauged Skyrme model on 
$\R_3$ is the selection of a suitable topological charge density which could 
be exploited to give a lower bound on the energy density of a suitably 
modified version of the $3$ dimensional Skyrme model. Since we are 
dealing with a Skyrme-Sigma model featuring a constrained field and a 
global $O(4)$ invariance, we expect that the relevant topological charge 
density will be only a {\it locally} total derivative quantity as was realised 
in \cite{Schroers}. The topological charge density of the $O(4)$ 
Skyrme\cite{Skyrme} model is given by

\begin{equation}
\label{2.1}
\varrho_0 =\varepsilon_{ijk} \varepsilon^{abcd} \partial_i \phi^a 
\partial_j \phi^b \partial_k \phi^c \phi^d
\end{equation}
where the index $a=\alpha  , 4 $, and $\alpha  =1,2,3$, etc., and the volume 
integral of \r{2.1} yields the winding number provided that the fields 
exhibit the appropriate asymptotic behaviour.

The gauged version of the density \r{2.1} , namely
\begin{equation}
\label{2.2}
\varrho_1 =\varepsilon_{ijk} \varepsilon^{abcd} D_i \phi^a D_j \phi^b D_k 
\phi^c \phi^d.
\end{equation}
is given in terms of the $d=3$ version of the covariant derivative \r{1.2}
\begin{equation}
\label{2.3}
D_i\phi^{\alpha} =\partial_i \phi^{\alpha} +\varepsilon^{\alpha \beta 
\gamma} A^{\beta}_i \phi^{\gamma}, \qquad \qquad D_i\phi^4 =\partial_i 
\phi^4 .
\end{equation}
The two densities \r{2.1} and \r{2.2} are related as follows
\begin{equation}
\label{2.4}
\varrho_1 =\varrho_0 +3\varepsilon_{ijk} \phi^4 F^{\alpha}_{ij} 
D_k\phi^{\alpha} +\partial_i \Omega_i
\end{equation}
where $F^{\alpha}_{ij}$ is the $SO(3)$ curvature of the connection defined 
by the first member of \r{2.3}, and the density $\Omega_i$ is given by
\begin{equation}
\label{2.5}
\Omega_i =3\varepsilon_{ijk} \phi^{\alpha} (2A^{\alpha}_j \partial_k 
\phi^4 -\phi^4 F^{\alpha}_{jk}).
\end{equation}
The latter is a {\it gauge variant} quantity, like the density $\varrho$, 
\r{2.1}. As we shall see below, the topological charge density will turn out 
to be
\begin{equation}
\label{2.9}
\varrho =\varrho_0 +\partial_i \Omega_i = \varrho_1 -3\varepsilon_{ijk} 
\phi^4 F_{ij}^{\alpha} D_k \phi^{\alpha},
\end{equation}
which is manifestly {\it gauge invariant} as it should be.

The volume integral of the topological charge density $\varrho$ of \r{2.9}  
in fact reduces to the volume integral of the winding number density 
$\varrho_0$ of \r{2.1}, provided that the surface integral of $\Omega_i$ 
vanishes. This will be verified below for the spherically symmetric field 
configurations. Here we procede to explain our criteria by means of stating 
the following Belavin inequality:

\begin{equation}
\label{2.6}
(\lambda_1 D_i\phi^a -\frac{\kappa_2}{2} \varepsilon_{ijk} 
\varepsilon^{abcd} D_{[j}\phi^b D_{k]}\phi^c \phi^d)^2 \ge 0
\end{equation}
where the square brackets$[..]$ on the indices imply antisymmetrisation, 
the constant $\kappa_2$ has the dimension of length and the constant 
$\lambda$ is dimensionless. The inequality \r{2.6} is the first member of 
the two pairs, for the second member of which there are two options. These 
are, respectively,

\begin{equation}
\label{2.7a}
(\kappa_0 F^{\alpha}_{ij} +\frac{1}{2} g(\phi^4) \varepsilon_{ijk} 
D_k\phi^{\alpha})^2 \ge 0
\end{equation}
\begin{equation}
\label{2.7b}
(\kappa_0 h(\phi^4)F^{\alpha}_{ij} +\frac{1}{2} \varepsilon_{ijk} 
D_k\phi^{\alpha})^2 \ge 0,
\end{equation}

\noindent
in both of which the constant $\kappa_0$ has the dimension of length and 
the yet undetermined functions $g$ and $h$ depend only on $\phi^4$, 
which according to the second member of \r{2.2} is the gauge invariant 
component of $\phi^a$.

As usual, the energy density will be the sum of the square terms of \r{2.6} 
and \r{2.7a}, and respectively \r{2.7b}, when these are expanded.  Choosing 
the functions $g(\phi^4)$ and $h(\phi^4)$ appropriately in each case, 
results in the cross term coinciding with the topological charge density 
$\varrho$ defined by \r{2.9}. The topological charge, which is the volume 
integral of the latter, will be equal to the volume integral of $\varrho_0$ 
since the surface integral of $\Omega_i$ vanishes as will be verified 
explicitly in the case of the spherically symmetric field configuration 
below. It then follows that the lower bound on the energy is the degree of 
the map, namely the winding number of the Hedgehog given by the 
volume integral of $\varrho_0$, {\it provided} that suitable asymptotic 
conditions are satisfied. These are stated as usual to be

\begin{equation}
\label{2.8}
\lim_{|\vec x| \rightarrow 0}\phi^4 = -1,\:\:\:\:\:\:\lim_{|\vec x| 
\rightarrow 
\infty}\phi^4 = 1.
\end{equation}

The appropriate choices for the functions $g$ and $h$ turn out to be the 
same, and namely,
\[
g(\phi^4)=h(\phi^4)=3\lambda_1 (\frac{\kappa_2}{\kappa_0})\phi^4
\]
respectively, leading to the two alternative Hamiltonian densities given by

\begin{equation}
\label{2.10a}
{\cal H}_1 = \kappa_0^2 (F^{\alpha}_{ij})^2+\frac{9\lambda_1^2}{2} 
(\frac{\kappa_2}{\kappa_0})^2 (\phi^4)^2 (D_i\phi^{\alpha})^2 + 
\frac{\lambda_1^2}{2} (D_i\phi^a)^2 + \frac{1}{2}\kappa_2^2 (D_{[i}\phi^a 
D_{j]}\phi^b )^2
\end{equation}
and
\begin{equation}
\label{2.10b}
{\cal H}_2 = 9\lambda_1^2 \kappa_2^2 (\phi^4)^2 
(F^{\alpha}_{ij})^2+\frac{1}{2} (D_i\phi^{\alpha})^2 + 
\frac{\lambda_1^2}{2} (D_i\phi^a)^2 + \frac{1}{2}\kappa_2^2 (D_{[i}\phi^a 
D_{j]}\phi^b )^2.
\end{equation}

\noindent
We notice that both Hamiltonian densities break the global $O(4)$ 
symmetry of the corresponding ungauged sigma model, namely the Skyrme 
model. This is manifested through the appearance of $|\phi^{\alpha}|^2$ 
and $|D_i\phi^{\alpha}|^2$ instead of $|\phi^a|^2$ and $|D_i\phi^a|^2$ 
respectively. This situation can easily be altered by adding suitable positive 
definite terms to each of \r{2.10a} and \r{2.10b} without invalidating the 
respective topological inequalities. In the case of \r{2.10a} for example, the 
quantity to be added is $\frac{9\lambda_1^2 \kappa_2^2}{2\kappa_0^2}$ 
times $(\phi^{\beta})^2 (D_i\phi^{\alpha})^2 +(\partial_i \phi^4)^2$. Both 
\r{2.10a} and \r{2.10b} then take the form

\begin{equation}
\label{2.10c}
{\cal H}_0 = \eta_0^2 (F^{\alpha}_{ij})^2 + \frac{\tau_1^2}{2} (D_i\phi^a)^2 
+ \frac{\eta_2^2}{8} (D_{[i}\phi^a D_{j]}\phi^b )^2,
\end{equation}
in which the constants $\eta_0$ and $\eta_2$ have the dimensions of 
length, and the constant $\tau_1$ is dimensionless. The Hamiltonian 
density \r{2.10c} coincides with the system proposed by Fadde'ev in 
Ref.\cite{Fadde'ev}.

\noindent
The volume integrals of \r{2.10a}, \r{2.10b} and \r{2.10c} are bounded from 
below by the winding number according to

\begin{equation}
\label{2.11}
\int d^3 x {\cal H} \ge \int d^3 x \varrho =\int d^3 x \varrho_0 ,
\end{equation}
provided that the field configurations satisfy at least the asymptotic 
conditions \r{2.8}.

It is not possible to saturate the inequalites \r{2.11} by saturating the 
inequalities \r{2.6} and \r{2.7a} separately to minimise {\it absolutely} the 
energy corresponding to \r{2.10a}, and, \r{2.6} and \r{2.7b} separately to 
minimise the energy of \r{2.10b} {\it absolutely}. By minimising absolutely 
we mean solving the system by some first order Bogomol'nyi equations, and 
in the case of \r{2.10c} this is not possible even in principle. For the models 
given by \r{2.10a} and \r{2.10b}, while in principle possible, the saturated 
versions of \r{2.6} and \r{2.7a}, \r{2.7b} respectively, are overdetermined 
as is the case also for the (ungauged) Skyrme model\cite{Skyrme}. This 
contrasts with the corresponding situation for the $2$ dimensional $O(3)$ 
model where both the ungauged\cite{Belavin} and the 
gauged\cite{Schroers} models support self-dual solutions. In the following 
therefore, we are concerned only with solutions of the second order Euler-
Lagrange equations of the models \r{2.10a}, \r{2.10b} and \r{2.10c}, and not 
with the solutions of some first order Bogomol'nyi equations.

\section{\bf Spherically Symmetric Fields}

The spherically symmetric fields are given by the Ansatz
\begin{equation}
\label{3.1}
\phi^{\alpha} =\hat x_i\: \: \sin f(r),\qquad \qquad \phi^4 =\cos f(r)
\end{equation}
\begin{equation}
\label{3.2}
A^{\alpha}_i =\frac{a(r)-1}{r} \varepsilon_{i\alpha \beta} \hat x_{\beta} 
\end{equation}
yielding the following field strengths
\begin{equation}
\label{3.3}
D_i\phi^{\alpha} =\frac{a\: \sin f}{r} \delta^{\alpha}_{i} +(f'\: \cos f 
-\frac{a\: \sin f}{r}) \hat x_i \hat x^{\alpha}
\end{equation}
\begin{equation}
\label{3.4}
F^{\alpha}_{ij}=\frac{a'}{r} \varepsilon_{ij\alpha} -\left (\frac{a'}{r} 
-\frac{a^2 -1}{r^2} \right) \varepsilon_{ij\beta} \hat x_{\beta} \hat 
x_{\alpha}.
\end{equation}
In \r{3.1} -\r{3.4} $\hat x$ is the unit position vector and $f'=df/dr$ etc.

As stated in the previous section, we are concerned exclusively with the 
second order Euler-Lagrange equations here, and not in first order 
Bogomol'nyi equations obtained by saturating \r{2.6} and \r{2.7a}, and 
respectively \r{2.6} and \r{2.7b}. These are easily verified in this 
spherically symmetric configuration \r{3.1} and \r{3.2} to be 
overdetermined. We define the static Hamiltonian density of the one 
dimensional subsystem obtained by substituting \r{3.3} and \r{3.4} into 
\r{2.10a}, \r{2.10b} and \r{2.10c} respectively, by
\begin{equation}
\label{3.5}
\int d\rho H[f,f_{\rho} ;a,a_{\rho}] = \frac{1}{4\pi} \int d^3 x {\cal H}
\end{equation}
where the dimensionless radial variable $\rho$ is defined in the two cases 
as $\rho =\frac{r}{\kappa_0}$ and $\rho =\frac{r}{\kappa_2}$ 
respectively, and $f_{\rho} =\frac{df}{d\rho}$. The definition \r{3.5} is 
made, up to an unimportant constant multiple in each case.

Since the alternative models \r{2.10a}, \r{2.10b} and \r{2.10c} are 
qualitatively similar, we shall restrict ourselves in the following to the 
detailed asymptotic and numerical study of one of these models only. We 
find it more natural to prefer models \r{2.10a} and \r{2.10b} to \r{2.10c} 
because the former satisfy the minimal topological inequality \r{2.11}, 
albeit without saturating it, while model \r{2.10c} satisfies an inequality 
derived from \r{2.11} itself. Next, we eschew model \r{2.10b} because of its 
unconventional Yang-Mills term. Thus we restrict our considerations 
below to the model given by \r{2.10a}.

In terms of the dimensionless parameter $\lambda_2 
=(\frac{\kappa_2}{\kappa_0})$, the resulting one dimensional Hamiltonian 
density for the models given by \r{2.10a} is

\[H_1=2[2a_{\rho}^2 +\frac{(a^2 -1)^2}{\rho^2} ]+\frac{9\lambda_1^2 
\lambda_2^2}{2}  \cos^2 f [\rho^2 f_{\rho}^2 \cos^2 f +2a^2 \sin^2 f] \]
\begin{equation}
\label{3.6}
+ \frac{\lambda_1^2}{2} [\rho^2 f_{\rho}^2 + 2a^2\sin^2 f ]+2\lambda_2^2 
a^2 \sin^2 f [2f_{\rho}^2 +\frac{a^2\sin^2 f}{\rho^2}]
\end{equation}

We first check that the topological charge, namely the volume integral of 
$\varrho$ given by \r{2.9} reduces to the usual winding number, namely 
the volume integral of $\varrho_0$ for the spherically symmetric field 
configuration \r{3.1} and \r{3.2} when the appropriate asymptotic 
conditions for the function $f(r)$
\begin{equation}
\label{3.7}
\lim_{r \rightarrow \infty} f(r) = 0,\qquad \qquad \lim_{r \rightarrow 0} 
f(r) =\pi
\end{equation}
are satisfied. In that case the volume integral of $\varrho_0$ is guaranteed 
to be the {\it unit} topological charge of the Hedgehog.

Concerning the asymptotic behaviour required of the fuction $a(r)$ in the 
region $r\ll 1$, this is determined by regularity at the origin, while in the 
region $r\gg 1$ there are several possibilities consistent with a power 
decay of the function $a(r)$ at infinity, which is a necessary condition for 
{\it finite energy} solutions. These asymptotic values are $a(\infty)=\pm 1$ 
and $a(\infty)=0$. Unlike in $SU(2)$ Higgs theory\cite{'tHooft-Polyakov}, 
the behaviour of the gauge field and hence of the function $a(r)$ at 
infinity is not directly relevant to the topological stability of the soliton. In 
the latter case\cite{'tHooft-Polyakov} the topological charge coincides with 
the magnetic flux of the monoplole field, while here the topological charge 
is the degree of the map, which is the unit winding number for the 
spherically symmetric fields under consideration. Using the magnetic flux 
density $B_i =\frac{1}{2}\varepsilon_{ijk} \phi^{\alpha}F_{jk}^{\alpha}$, 
for the spherically symmetric field configuration, we calculate the surface 
integral for the magnetic flux
\begin{equation}
\label{3.8}
\Phi =\int \vec B . d\vec S=4\pi [(a^2 -1) \sin f ]|_{r=\infty}.
\end{equation}
Irrespective of the value of $a(\infty)$, the flux $\Phi$ vanishes by virtue 
of the second member of \r{3.7}. The models at hand therefore {\it do not} 
describe magnetic monopoles.

Before stating our asymptotic conditions for the function $a(r)$, we note 
that for the topological charge density $\varrho$ given by \r{2.9} to 
reduce to the usual winding number density $\varrho_0$ , the surface 
integral of $\Omega_i$ must vanish. This is seen by calculating the 
relevant one dimensional integrand, namely the quantity $r^2 \hat x_i 
\Omega_i$, from its definition \r{2.5},
\begin{equation}
\label{3.9}
r^2 \hat x_i \Omega_i =(a^2 -1)\sin 2f
\end{equation}
which on the infinite 2-sphere clearly vanishes irrespective of which of 
the asymptotic values $a(\infty)=\pm 1$ or $a(r)=0$ holds, provided that 
the corresponding condition stated by the second member of \r{3.7} does 
hold.

Anticipating the results of our numerical integration to be carried out 
below, we state the asymptotic values of $a(r)$ as
\begin{equation}
\label{3.10}
\lim_{r \rightarrow 0} a(r)=1, \qquad \qquad \lim_{r \rightarrow \infty} 
a(r) =0.
\end{equation}

The Euler-Lagrange equations for the model given by \r{3.6}, with respect 
to the arbitrary variations of the functions $f(r)$ and $a(r)$ are 
respectively,

\[
\lambda_1^2 (1+9\lambda_2^2 \cos^4 f)(\rho^2 f_{\rho})_{\rho} 
+8\lambda_2^2 \sin^2 f \: (a^2 f_{\rho})_{\rho}
\]
\[
-18\lambda_1^2 \lambda_2^2 \sin f \cos f \: [\rho^2 f_{\rho}^2 \cos^2 f +a^2 
(\cos^2 f -\sin^2 f)]
\]
\begin{equation}
\label{3.11}
+8\lambda_2^2 a^2 \sin f \cos f \: [f_{\rho}^2 -\frac{a^2 \sin^2 f}{\rho^2}] 
-2\lambda_1^2 a^2 \sin f \cos f =0
\end{equation}
and
\begin{equation}
\label{3.12}
a_{\rho \rho}-\frac{a}{\rho^2} (a^2 -1)- a \sin^2 f \: [\frac{9}{4} 
\lambda_1^2 \lambda_2^2 \cos^2 f +\frac{1}{4} \lambda_1^2 +\lambda_2^2 
(f_{\rho}^2+\frac{a^2 \sin^2 f }{\rho^2} )]=0,
\end{equation}

The asymptotic values in the $r\ll 1$ region are given by the first members 
of \r{3.7} and \r{3.10}, and for the model \r{2.10a} we find the following 
behaviours
\begin{equation}
\label{3.15}
f(\rho)=\pi +A\rho + o(\rho^3)
\end{equation}
\begin{equation}
\label{3.16}
a(\rho)=1+B\rho^2 + o(\rho^4),
\end{equation}
which lead, as expected, to differentiable fields \r{3.3} and \r{3.2} at the 
origin.

The asymptotic behaviours of the solution in the $r\gg 1$ region are also 
power decays like for the usual (ungauged) Skyrmion\cite{Skyrme}. 
(Exponential decay can be obtained by incorporating a suitable gauge 
invariant $O(4)$ breaking potential\cite{Roche}.) Since we shall integrate 
the field equations with the asymptotic conditions \r{3.7} and \r{3.10}, we 
give the corresponding asymptotic solutions in this region
\begin{equation}
\label{3.17}
f(\rho)=\frac{C}{\rho} ,\qquad a(\rho)=\frac{D}{\rho^{\beta}}
\end{equation}
in which $\beta =\frac{1}{2} [\sqrt {(C^2 \lambda_1^2(\lambda_1^2 
+9\lambda_2^2)-3)} -1]$, and the constants $C$ and $D$ will not be 
computed.

Having solved the relevant Euler-Lagrange equations \r{3.10} and \r{3.11} 
in the asymptotic regions $\rho <<1$ and $\rho >>1$, we procede to integrate 
them numerically, subject to the asymptotic conditions \r{3.2} and \r{3.3}. 
The constants $A$ and $B$ in \r{3.15} and \r{3.16} are fixed by the 
numerical integration.

The numerical integrations have been performed for the values of the 
dimensionless coupling constants $\lambda_1 =\lambda_2 =1.5$, and 
$\lambda_1 =\lambda_2 =1.4$. The values of the pair of constants $\{ A,B\}$ 
were fixed in each of these cases respectibely to be
\[
\{ A,B\} =\{-1.9606513707554, -5.6289634247230\} 
\]
\[
\{ A,B\} =\{-1.8663921477326, -4.7278588179841\} 
\]
The profiles of the function $f(\rho)$ are given in Figure 1, and the 
profiles of the function $a(\rho)$ in Figure 2. The profiles of the energy 
densities pertaining to each of these solutions are plotted in Figure 3, and 
they correspond to the total energies $E_1 =29.924879981245$ and $E_2 
=27.463710758882$ respectively.

\section{\bf Discussion and Summary }

Before proceding to summarise our results and making some qualitative 
comments, we give a breif quantitative description of the Baryonic current 
that the topological charge employed above pertains to. The latter is the 
volume integral of the density $\varrho$ given by \r{2.9}, which we 
identify with the fourth, time-like, component of this current $j^{\mu}$. 
Accordingly, the full Minkowskian vector current is

\begin{equation}
\label{4.1}
j^{\mu} =\varepsilon^{\mu \nu \rho \sigma} \varepsilon_{abcd} [D_{\nu} 
\phi^a D_{\rho} \phi^b D_{\sigma} \phi^c \phi^d -\frac{3}{4} F_{\rho 
\sigma}^{cd} D_{\nu} \phi^a \phi^b ].
\end{equation}

\noindent
The curvature field strength in \r{4.1} consists only of an $SU(2)$ field, say 
$F_{\mu \nu}^{\alpha \beta} =\varepsilon^{\alpha \beta \gamma} F_{\mu 
\nu}^{\gamma}$, with $F_{\mu \nu}^{\alpha 4} =0$. Accordingly, the 
second term in \r{4.1} can be re-expressed using

\begin{equation}
\label{4.2}
\varepsilon^{\mu \nu \rho \sigma} \varepsilon_{abcd} F_{\rho 
\sigma}^{cd} D_{\nu} \phi^a \phi^b =2\varepsilon^{\mu \nu \rho \sigma} 
[2F_{\rho \sigma}^{\alpha} D_{\nu}\phi^{\alpha} \phi^4 - \partial_{\nu} 
(F_{\rho \sigma}^{\alpha} \phi^{\alpha} \phi^4)].
\end{equation}

\noindent
The second term on the right hand side of \r{4.2} being a total divergence, 
its volume integral vanishes and hence can be neglected. It follows from 
\r{4.2} that the divergence of the current $j^{\mu}$ given by \r{4.1} is 

\begin{equation}
\label{4.3}
\partial_{\mu} j^{\mu} = -4\varepsilon^{\mu \nu \rho \sigma} 
\varepsilon^{\alpha \beta \gamma} D_{\mu}\phi^{\alpha} 
D_{\nu}\phi^{\beta} D_{\rho}\phi^{\gamma} \partial_{\sigma}\phi^4 .
\end{equation}

\noindent
The right hand side of \r{4.3} can be shown to be {\it locally total 
divergence} which means that its volume integral vanishes and hence can 
be ignored, leading to a conserved Baryonic current, $\partial_{\mu} 
j^{\mu} =0$. This is exactly what we expect, since the {\it vector} gauging 
\r{2.3} does not lead to the divergence of the current being equal to an 
anomaly. Our current \r{4.1} can be compared to that of Goldstone and 
Wilczek\cite{GW}, where the $O(4)$ sigma model has been gauged in the 
usual way according to $D_{\mu}\phi^a =\partial_{\mu} \phi^a + 
A_{\mu}^{[ab]} \phi^b$, and contrasted with the corresponding current of 
D'Hoker and Farhi\cite{DF} where the Skyrme model featuring the $SU(2)$ 
valied field $U$ has been gauged with the $(V-A)$ $SU(2)$ field according 
to $D_{\mu}U=\partial_{\mu} U-i\vec A_{\mu} . \vec \sigma U$. In the 
latter case\cite{DF}, the divergence of the Baryonic current equals the 
anomaly. 

We have unfortunately {\it not} succeeded to adapt the constructions 
employed in the $O(4)$ sigma model with field $\phi^a$, to the analogous 
case where the $SU(2)$ valued Skyrme field $U$ is used, which is the 
physically more interesting case as it leads to a non-conserved Baryonic 
current featuring the anomaly. Technically, this has come about because of 
our inability to reduce the topological charge density $\varsigma$ in this 
case, analogous to $\varrho$ used in the above, to the form

\begin{equation}
\label{4.4}
\varsigma =\varsigma_0 +\partial_i \tilde \Omega_i
\end{equation}

\noindent
in which $\varsigma_0 =\varepsilon_{ijk} TrU^{-1}\partial_i U \partial_j 
U^{-1} \partial_k U$ is equal to $\varrho_0$ defined by \r{2.1} with 
$U=\phi^a \tau^a$, $U^{-1} =\phi^a \tilde \tau^a$ defined at the end of 
Section 1. Had it turned out possible to establish \r{4.4}, then the topological 
charge would have been related to $\varsigma_1$, analogous to 
$\varrho_1$ in \r{2.2}, which would have been the time-like component of 
the topological Baryonic current

\begin{equation}
\label{4.5}
j^{\mu} =\varepsilon^{\mu \nu \rho \sigma} Tr[U^{-1} D_{\nu} U\: U^{-1} 
D_{\rho} U\: U^{-1} D_{\sigma} U +\frac{3}{4} U^{-1} F_{\rho \sigma} 
D_{\nu} U].
\end{equation}

\noindent
used in \cite{DF}, whose divergence $\partial_{\mu} j^{\mu}$ does not 
vanish but is equal to the chiral $SU_{\pm}$ anomaly. It would be very 
interesting if some other version of \r{4.4} could be found, which would 
lead to a lower bound on the static Hamiltonian.

To summarise, we have constructed three $SO(3)$ gauged versions of the 
$O(4)$ sigma model on $\R_3$, characterised by the static Hamiltonians 
\r{2.10a}, \r{2.10b} and \r{2.10c}. These are equivalent to the 
corresponding gauged versions of the Skyrme model\cite{Skyrme}. Models 
\r{2.10a} and \r{2.10b}have the feature of breaking the global $O(4)$ 
symmetry of the sigma model, but the latter can be restored by adding 
suitable positive definite terms to the Hamiltonian densities resulting in 
the model \r{2.10c} which was first proposed by Fadde'ev\cite{Fadde'ev}. 
The large $r$ asymptotic field configuration of the soliton presented here 
is $SO(3)$ symmetric and the magnetic flux of the corresponding field 
configuration vanishes. Accordingly this soliton model differs from the 
corresponding $SU(2)$ Higgs model\cite{'tHooft-Polyakov}, in which the 
asymptotic field configuration is $SO(2)$ symmetric and exhibits magnetic 
monopole flux.

These models admit finite energy topologically stable soliton solutions, 
whose energy is bounded from below by the winding number, which can 
be interpreted as the Baryon number. This topological bound is saturated 
by Bogomol'nyi equations, which however are overdetermined as in the 
case of the usual (ungauged) Skyrme model, and hence likewise our 
solitons are solutions to the full second order Euler-Lagrange equations. 
These were solved analytically only in the asymptotic regions $r<<1$ and 
$r>>1$, and the full integrations were performed numerically.

In the present work, we have restricted ourselves to the spherically 
symmetric case. As such the soliton in question carries Baryon number $1$. 
It would be interesting to find the Baryon number $2$ {\it axially 
symmetric} solutions, analogous to the corresponding axially symmetric 
solutions of the (ungauged) Skyrme model\cite{Manton}. Furthermore, 
since we know\cite{Roche} that for sigma models in odd dimensional spaces 
there are spherically symmetric solitons of arbitrary Baryon number $N$, 
it would be interesting to study these in the present model. These higher 
degree field configurations are characterised by their asymptotic values, 
which for small $r$ differ from \r{3.7} according to
\begin{equation}
\label{4.6}
\lim_{r \rightarrow \infty} f(r) = 0,\qquad \qquad \lim_{r \rightarrow 0} 
f(r) =N\pi .
\end{equation}
It would be interesting to integrate the Euler-Lagrange equations with the 
asymptotic conditions \r{4.6}, say with N=2, and to see whether the energy 
of that soliton is greater than twice the energy of the $N=1$ soliton, as is 
the case for the usual (ungauged) Skyrme model\cite{Jackson}. All these 
detailed questions are deferred to future investigations.

{\bf Acknowledgements} It is a pleasure to thank V.A. Rubakov for his 
patient explanations and for carefully reading the manuscript. We are 
grateful to B.J. Schroers for having brought Ref.\cite{Fadde'ev} to our 
attention. This work was supported in part by INTAS contract 93-1630.

\bigskip

\noindent
Figure Captions:

\noindent
Figure 1. Profiles of the function $f(\rho)$ for $\lambda =\lambda_1 
=\lambda_2 =1.5$ and $\lambda =1.4$. The higher curve pertains to 
$\lambda =1.5$.

\noindent
Figure 2. Profiles of the function $a(\rho)$ for $\lambda =\lambda_1 
=\lambda_2 =1.5$ and $\lambda =1.4$. The higher curve pertains to 
$\lambda =1.4$.

\noindent
Figure 3. Profiles of the energy densities corresponding to the solutions 
with $\lambda =\lambda_1 =\lambda_2 =1.5$ and $\lambda =1.4$. The higher 
curve pertains to $\lambda =1.5$.


\newpage


\begin{thebibliography}{99}

\bibitem{Fadde'ev} L.D. Fadde'ev, Lett. Math. Phys. {\bf 1} (1976) 289.
\bibitem{Witten} E. Witten, Nucl. Phys. {\bf 223} (1983) 433.
\bibitem{Rubakov} V.A. Rubakov, Nucl. Phys. {\bf 256} (1985) 509.
\bibitem{Skyrme}T.H.R. Skyrme, Proc. Roy. Soc.{\bf A260} (1961) 
127;\\Nucl.Phys. {31} (1962) 556.
\bibitem{Arthur} D.H. Tchrakian and K. Arthur, Phys. Lett. {\bf B352} 
(1995) 327.
\bibitem{Manvelyan} R.P. Manvelyan and D.H. Tchrakian, Phys. Lett. {\bf 
B352} (1995) 321.
\bibitem{Zakrzewski} see for example W.J. Zakrzewski, "Low dimensional 
sigma models", Adam Hilger, Bristol 1989,
\bibitem{Schroers} B.J. Schroers, Phys. Lett. {\bf 356} (1995) 291.
\bibitem{Roche} K. Arthur, G. Roche and D.H. Tchrakian, DIAS-STP-95-29
\bibitem{Belavin} A.A. Belavin and A.M. Polyakov, JETP Lett {\bf 22} (1975) 
245.
\bibitem{'tHooft-Polyakov} G. 't Hooft, Nucl. Phys. {\bf B79} (1974) 276; A.M. 
Polyakov, JETP Lett. {\bf 20} (1974) 194.
\bibitem{Manton} V.B. Kopeliovich and Shtern, Pisma v'ZhETF {\bf 45} 
(1987) 165; N.S. Manton, Phys. Lett. {\bf B192} (1987) 177; J. Verbraarschot, 
{\it ibid.} {\bf B195} (1987).
\bibitem{Jackson} A.D. Jackson and M Rho, Phys. Rev. Lett.{\bf 51} (1983) 
751.

\bibitem{GW} J. Goldstone and F. Wilczek, Phys. Rev. Lett. {\bf 47} (1981) 
986.

\bibitem{DF} E. D'Hoker and E. Farhi, Nucl. Phys. {\bf B241} (1984) 109.


\end{thebibliography}
\end{document}